**Effective Mass Ratio & positive colossal magnetoresistance of a Nano-wire**

Piyush Dua

Amity Institute for Advanced Research & Studies (Materials & Devices)
Amity University Uttar Pradesh, Sector-125, Noida-201301, India

<u>Abstract</u>

In the present work, a relation has been established between degree of polarization and effective mass ratio (EMR) and magnetoresistance (MR) of one-dimensional non-degenerate system (which can represent a nano-wire or a linear chain of atoms and molecules in one dimension) by using a non-degenerate Hubbard model, which includes diagonal and off-diagonal matrix elements of Coulomb interaction. EMR is one of the most important property, which provides information that how much itinerant the system is? Within the mean field approximation, it is found that, due to the presence of off-diagonal elements, the band narrowing effect dominates over the band splitting effect. The EMR varies with degree of polarization. EMR for majority (up spin) carriers decreases as magnetization increases below ferromagnetic transition temperature. MR decreases as system is cooled below the ferromagnetic transition temperature. A comparison of our results with existing experimental observations is made.

## 1. Introduction

The origin of ferromagnetism in one-dimensional itinerant electron systems remains one of the problems for which a lot of questions are yet to be answered [1-2]. Spintronic [3] phenomenon (which can be studied as a member of itinerant electron system) uses both the properties of electrons i.e. charge (for conductance) as well as spin (for memory). It is expected that utilization of spin-related phenomenon in information processing will extend the functionality of conventional devices based on new operating principles [3-5]. One of the requirements to calibrate spintronic devices is that the dimension should be very small. At reduced dimensionality, atomic nature dominates and Coulomb interactions play a crucial role in determining device properties, such as in nano-wires & Carbon nano-tubes [6, 7]. For the role of strong correlations in low dimensional conductors, see ref. [8]. It is well known that ultra small nano scale materials display many peculiar phenomenon not observed in macroscopic state and the study of nano scale materials has become the most interesting issue. For instance, 1D Co ferromagnetic nanowire was grown and magnetic anisotropy was measured and showed that in ferromagnetic state narrowing of 3d band takes place [1]. On the theoretical side, giant magnetic anisotropy energy was obtained in single Co atom and nanoparticles [9], magnetic anisotropy energy were calculated of Co nanochains [10], Delin et al [2], using density functional calculations showed that Pd nanowires exhibit magnetic ground state on reducing the dimension. Spin polarized transport in Cu [11] & Co nanowires [12] had been shown. Nilius et al [13] showed that artificial 1D Au chains show spin-polarized state as driven by effective mass reduction.

Komelj & group [14, 15] studied the effect of dimensionality for Fe & Co and found that (i) The spin moments increase only slightly with decreasing dimensionality and (ii) the orbital moments increase strongly with decreasing dimensionality when orbital polarization is taken into account. Nautiyal et al [16,17] have investigated the changes in electronic structure and magnetic moments of various transition metal elements and predicted the importance of spin-orbit interaction and electron correlations at reduced dimensions, Tang et al [18] studied the magnetic

& electronic properties of all 3d transition metal linear and zigzag nanowires and found that all are having stable or metastable ferromagnetic state.

The properties of these systems depend upon various internal & external parameters. Davila et al [19] suggested that magnetic anisotropy can be tuned as a function of external parameters, Komelj et al [20] studied the effect of substrate interaction and found that freestanding wires and wires at the substrate shows different magnetic anisotropy energy, Krawiec et al [21] studied the monoatomic gold wires on vicinal surfaces by using Hubbard model and predicted that wires strongly interact via surface with neighboring ones on the same terrace, Brovko et al [22] Demonstrated that atomic chains can enhance the exchange interaction between magnetic impurities, can change the sign of the exchange interaction or quench in completely by tailoring different nanostructures, Nielsen et al [23] studied the effect of electron-hole asymmetry on ferromagnetism in non-magnetic doped semiconductors such as $Ga_{1-x}Mn_xAs$ using Hubbard & t-J models, Ataca et al [24] studied thoroughly the light transition elements monoatomic chains and found that geometric structure influence strongly the electronic and magnetic properties of the chains. Choi et al [25] suggested that Cs-Au chain may be the best candidate for a linear atomic chain because the band gap can be tuned by changing the template and substrate.

As far as the conducting devices are concerned, the effective mass ratio (EMR) always considered to being an important quantity. EMR is one of the properties, which provides information that how much itinerant the system is? EMR depends upon various parameters such as type of the system, degree of doping, interaction parameters and surrounding environment.

The recently measured, Hall-effect data suggests that simultaneously with the on-set of magnetic order the effective charge-carrier (up spin electrons) concentration increases significantly. It has been observed that the effective mass of the itinerant charge-carriers decreases with increasing external magnetic field and decreasing temperature [26-28]. Kierspe *et al* [29] observed the electrical resistivity ρ of the three ferromagnetic metals (Fe, Ni and Co) and an anomalous drop of ρ below the ferromagnetic transition temperature $T_c$ has been shown. Spectral weight (for up spin electrons) increases on decreasing the temperature below $T_c$ or on increasing the external magnetic filed. This feature has been experimentally observed in various compounds like in manganites, hexaborides and $Ga_{1-x}Mn_xAs$ [29].

In the present work, we have established a relation between EMR and degree of polarization. We have considered a single-band non-degenerate Hubbard like tight binding Hamiltonian to represent a nanowire (a 1-dimensional continuous chain of atoms). The Hamiltonian includes, both the diagonal & off-diagonal elements of Coulomb interaction [6].

The effect of these parameters on the degree of polarization is discussed elsewhere [30-32]. Here we will discuss the effect of degree of polarization on EMR in context to nanowire. In Section 2, the detail of Hamiltonian is discussed and an expression is derived between EMR & degree of polarization. In section 3, we present the results and discussions. The section 4 contains the concluding remarks & future plans.

## 2. Details of Hamiltonian

If one considers, a linear chain of an itinerant electron system as in figure 1. Figure 1(a) shows a linear chain of similar atoms like Fe, figure 1(b) represent a linear chain of 'AB' type compound like MnAs and figure 1(c) represents a linear chain of 'AB' type compound doped with 'C' atom and 'C' atom replaces a 'A' atom at very few places becoming dilute doped compound $A_{1-x}C_xB$ like $Ga_{1-x}Mn_xAs$. The systems shown in figure 1(a) & (b) can be undoubtly, represented by Hubbard model. Assuming that on replacing 'A' by 'C' in figure 1(c) doesn't effect the parameters of Hubbard model to great extent, then figure 1(c) can also be represented by Hubbard model.

On the other hand, a linear chain of atoms and molecules can be represented by one-dimensional, non-degenerate extended Hubbard model, which contained diagonal and off-diagonal elements of Coulomb interaction. As a first approximation, the Vander-wall interaction is not included in the Hamiltonian, which becomes important at low dimensions.

$$H = -t \sum_{\langle ij \rangle \sigma} (c_{i\sigma}^+ c_{j\sigma} + H.C.) + U \sum_i n_{i\uparrow} n_{i\downarrow} + V \sum_{\langle ij \rangle} n_i n_j + J \sum_{\langle ij \rangle \sigma \sigma'} c_{i\sigma}^+ c_{j\sigma'}^+ c_{i\sigma} c_{j\sigma}$$

$$+ P \sum_{\langle ij \rangle} (c_{i\uparrow}^+ c_{i\downarrow}^+ c_{j\downarrow} c_{j\uparrow} + H.C.) + K \sum_{\langle ij \rangle \sigma \sigma'} c_{i\sigma}^+ c_{i\sigma}^+ c_{j\sigma} c_{i\sigma} \qquad (1)$$

Here t is hoping parameter, U is the intra-site Coulomb interaction, V is the inter-site Coulomb interaction, J is the exchange interaction, P is the pair-hopping interaction and K is the hybrid interaction (correlated hopping), between nearest neighbor sites. $c_{i\sigma}^+ (c_{i\sigma})$ creates (annihilates) an electron of spin $\sigma$ at site i. $n_i = n_{i\uparrow} + n_{i\downarrow}$ is total number of electron at site i. H.C. is the hermitian conjugate. The bracket $\langle ij \rangle$ restricts the sum to the nearest-neighbor (NN) sites i, j. The operator $n_{i\sigma} = c_{i\sigma}^+ c_{j\sigma}$ measures the occupancy of the site i with electrons of spin $\sigma$. The hybrid interaction takes care of the electron-hole asymmetry as well as the fluctuations in one dimension and has been found, within the mean-field approximation, more effective than all other inter-site interactions [30].

## 3. Results and discussions

We have chosen a quarter filled band using the constant density of states, with chemical potential μ chosen such that the total number of electrons comes out to be 0.5 (i.e. band is quarter filled). It will be discussed later in this study, that the quarter filling is very crucial.

Using Green's function equation of motion technique [6] and within mean field approximation, after a few steps of algebra, one gets the expression for EMR for non-degenerate system [30]

$$\frac{m_\sigma^*}{m} = \frac{D}{D_\sigma^*} = (1 - 2j_1 I_\sigma - 2j_2 I_{-\sigma} - 2k_2 n_{-\sigma})^{-1} \tag{2}$$

Here m and D are the spin-independent bare mass and bandwidth, m* & D* are effective mass & bandwidth and $I_\sigma$ is bond-charge. In terms of density of states ρ(ε), $I_\sigma$ may be written as

$$I_\sigma = \int_{-D/2}^{D/2} \rho(\varepsilon) \left(-\frac{\varepsilon}{D/2}\right) f[E_\sigma(\varepsilon)] d\varepsilon \tag{3}$$

For square density of states

$$\rho(\varepsilon) = \frac{1}{D} \quad \text{for} \quad -D/2 \le \varepsilon \le D/2 \tag{4}$$

$$E_\sigma(\varepsilon_k) = (1 - 2j_1 I_\sigma - 2j_2 I_{-\sigma} - 2k_2 n_{-\sigma})\varepsilon_k - \frac{\sigma D k_1}{2} M + D k_2 I_{-\sigma} - \mu \tag{5}$$

$$j_1 = \frac{zJ}{D} - \frac{zV}{D} \tag{6a}$$

$$j_2 = \frac{zJ}{D} + \frac{zP}{D} \tag{6b}$$

$$k_1 = \frac{U}{D} + \frac{zJ}{D} \tag{6c}$$

$$k_2 = \frac{zK}{D} \tag{6d}$$

In eq. 5, $\mu$ is the chemical potential, first term on right hand side is showing the band narrowing on the introduction of off-diagonal matrix elements of Coulomb interaction, whereas, second term is showing usual band-splitting phenomenon.

The optical (spectral) weight W (T, H) may be calculated as

$$W(T, H) = \frac{2}{\pi e^2} \int_0^{\omega_m} \sigma_1(\omega) \, d\omega = \frac{n}{m^*} \qquad (7)$$

where $\sigma_1(\omega)$ is the optical conductivity.

Under the assumption of a constant relaxation time, the magnetoresistance may be estimated from the change in effective mass with spin polarization

$$\frac{\Delta\rho}{\rho} = \frac{\rho(H) - \rho(0)}{\rho(0)} = \left[ \frac{\sum_\sigma n_\sigma(H)\{1 - 2j_1 I_\sigma(H) - 2j_2 I_{-\sigma}(H) - 2k_2 n_{-\sigma}(H)\}}{\sum_\sigma n_\sigma(0)\{1 - 2j_1 I_\sigma(0) - 2j_2 I_{-\sigma}(0) - 2k_2 n_{-\sigma}(0)\}} - 1 \right] \qquad (8)$$

Fig. 2 shows behavior of saturation magnetization $M_s$ as a function of band filling n. It is clear from fig. 2 that magnetization is least favored around half band filling. For $j_2 = 0.2$, we observe quite surprising results. For $k_2 = 0.6$, the magnetization M = n upto $n \cong 0.55$ and drops rapidly to zero for $n > 0.55$. For higher values of $k_2$ this drop (of magnetization from its saturation value to zero) takes place at higher values of band filling n. For $k_2 = 0.9$, the drop takes place around n = 1.0. Similar behavior of magnetization as a function of d-band filling has been observed experimentally in the itinerant ferromagnetic systems $Fe_{1-x}Co_xS_2$ and $Co_{1-x}Ni_xS_2$ [33]. In our study of degenerate Hubbard model of itinerant ferromagnetism [34], one peculiar behavior can be seen, that at (or near) quarter filling and three fourth filling, the required values of interaction parameters (for on-set and saturation magnetization) is independent of variation in inter-site coulomb interaction, so we can see the effect of correlated hoping, specifically. That's why we have chosen quarter band filling for our present study.

Fig. 3(a) & (b) show the variation of EMR as a function of temperature for different values of interaction parameters. Above $T_c$ the EMR is same for both up & down spin electrons and it is more than 1, i.e. even in paramagnetic state electron can't conduct as free electron. But as temperature reduces, below $T_c$, the degree of polarization increases & at absolute zero system is fully polarized. It is clear from the fig. 3, that as the degree of polarization increases for up spin electron, the EMR decreases whereas for down spin electrons the EMR increases sharply, below $T_c$. This feature of effective mass reduction has already been reported in $Ga_{1-x}Mn_xAs$ [28], which is a dilute magnetic semiconductor. This property has been reported by Sarma et al [35] and given a name 'Spintronic Effective Mass'. They had reported that EMR not only depends upon density but also on degree of polarization in 2 & 3 dimensions, within the leading order single loop self energy expansion in the dynamically screened Coulomb interaction. Hirsch [36] proposed that effective mass reduction would be the silent feature of itinerant ferromagnetism.

Fig. 4 shows behavior of optical weights for spin $\sigma$ electrons as a function of temperature. Spectral weight of up spin electrons increases as magnetization increases below $T_c$ and for down spin electrons, it decreases. Singley *et al* [28] have recently performed infrared spectroscopic measurements on itinerant ferromagnet $Ga_{1-x}Mn_xAs$. Their results of spectral weights, for ferromagnetic sample, resemble qualitatively with our results of total spectral weight for h = 0 (i.e., in absence of applied magnetic field).

Fig. 5 shows the behavior of magnetoresistance as a function of temperature for different magnetic fields. It is clear that the magnetoresistivity is maximum at $T_c$ and as the system is cooled below $T_c$, the magnetoresistivity decreases. The change is as large as 30-40%, which comes in the colossal magnetoresistance category. The band narrowing also modifies the magnetoresistance in a doped magnetic semiconductor [37] and a change of resistance of about 30% is found in Fe1-xCoxSb2, which comes in the category of positive colossal magnetoresistance.

## 4. Conclusions

Spintronics devices that utilize the spin degree of freedom of a charge to store, process or transmit information, may be better performers than their traditional electronic counterparts, if special properties of spin are exploited in the design. In the last few years, spintronics has emerged as a challenging and fast growing field of research. Spintronic device may overtake the conventional electronic devices, where it is not the electron charge but the electron spin that carries information. While conduction, conducting electrons lose their spin state (memory) through collisions with phonons, surrounding electrons and impurities. To overcome this effect, $T_1$ can be enlarged by reducing the dimension or by using the lighter elements of the periodic table [38]. Another important aspect is that the EMR in ferromagnetic state should become less than the paramagnetic state so that the electrons become more mobile in ferromagnetic state. With the help of this effect one can get a good competitive device. It has been noticed in 1D Au chains that in reduce dimension EMR is less that the bulk [13]. Effective mass reduction is related to band narrowing phenomenon, which is reflected by eq. (2). Band narrowing has been observed in 1D monatomic metal (Co) chains [1]. The effect of band narrowing has been seen on the magnetoresistance in a doped nearly magnetic semiconductor [37] and found positive colossal magnetoresistance.

In this work, we have proposed that 1D chain of (Fe, Co or Ni.) atoms can be represented by extended Hubbard model. Within the mean field approximation, we have found that EMR decreases in ferromagnetic state as temperature decreases as well as spectral weight for up spin electrons increases, as magnetization increases below $T_c$. Our results are in good qualitative agreement with observed experimental behavior of $Ga_{1-x}Mn_xAs$ [28]. An effort has been made to make a bridge between the present works with the emerging technology based on spintronic phenomenon. The work done here is the first step to understand the physics of one-dimensional chains using Hubbard model in presence of off-diagonal matrix elements of Coulomb interaction. Next step would be to include spin-orbit and Vander-wall interactions.

**Acknowledgement:** This work is sponsored by Board of Research in Nuclear Sciences (Grant No. 2006/37/43/BRNS/2289), Department of Atomic Energy (DAE), Government of India. I am very grateful to Dr. Ashok K. Chauhan, Founder President, Ritanand Balved Education Foundation for providing me necessary facilities for present study. I am very much thankful to Dr. M. Lavanya, Scietific Officer 'G', Bhabha Atomic Research Centre for continuous support and guidance for the present work.

**References**

[1]  Gambardella P, Dallmeyer A, Maiti K, Malagoll M C, Eberhardt W, Kern K and Carbone C 2002 Nature **416** 301

[2]  Delin E, Tosatti E and Weht R 2004 Phys. Rev. Lett. **92** 057201

[3]  Zutic I, Fabian J and Sarma S Das 2004 Rev. Mod. Phys. **76** 323

[4]  Wolf S A, Awschalom D D, Buhrman R A, Daughton J M, Molnar S von, Roukes M L, Chtchelkanova A Y and Treger D M 2001 Scinece **294** 1488

[5]  Prinz G 1995 Phys. Today **48(4)** 58

[6]  Hubbard J 1963 Proc. R. Soc. London Ser **276A** 238

[7]  Tutuc E, Melinte S and Shayegan M 2002 Phys. Rev. Lett. **88** 036805

[8]  Tremblay A-M S, Bourbonnais C and Senechal D Cond-mat **/ 0005111**.

[9]  Gambardella P, Rusponi S, Veronese M, Dhesi S S, Grazioli C, Dallmeyer A, Cabria I, Zeller R, Dederichs P H, Kern K, Carbone C and Brune H 2003 Science **300** 1130

[10] Hong J and Wu R Q 2004 Phys. Rev. B **70** 060406

[11] Gillingham D M, Muller C and Bland J A C 2004 J. Appl. Phys. **95** 6995

[12] Rodrigues V, Bettini J, Silva P C and Ugarte D 2003 Phys. Rev. Lett. **91** 096801

[13] Nilius N, Wallis T M and Ho W 2002 Science **297** 1853

[14] Ederer C, Komelj M and Fahnle M 2003 Phys. Rev. B **68** 052402

[15] Komelj M, Ederer C, Davenport J W and Fahnle M 2002 Phys. Rev. B **66** 140407

[16] Nautiyal T, Rho T H and Kim K S 2004 Phys. Rev. B **69** 193404

[17] Nautiyal T, Youn S J and Kim K S 2003 Phys. Rev. B **68** 033407

[18] Tang J C and Guo G Y 2007 Phys. Rev. B **76** 094413

[19] Davila J D and Pastor G M 1998 Phys. Rev. Lett. **81** 208

[20] Komelj M, Steiauf D and Fahnle M Cond-mat / **0601445**

[21] Krawiec M, Kwapinski T and Jalochowski M 2005 Phys. Stat. Sol. (b) **242** 332

[22] Brovko O O, Ignatiev P A, Stepanyuk V S and Bruno P, Cond-mat / **0804.3298**

[23] Nielsen E and Bhatt R N Cond-mat / **0705.2038**.

[24] Ataca C, Cahangirov S, Durgun E, Jang Y –R and Ciraci S, Cond-mat / **0801.1178**

[25] Choi Y C, Lee H M, Kim W Y, Kwon S K, Nautiyal T, Cheng D -Y, Vishwanathan K and Kim K S 2007 Phys. Rev. Lett. **98** 076101

[26] Broderick S, Ruzicka B, Degiorgi L and Ott H R 2002 Phys. Rev. B **65** R121102

[27] Paschen S, Pushin D, Schlatter M, Vonlanthen P, Ott H R, Young D P and Fisk Z 2000 Phys. Rev. B **61** 4174

[28] Singley E J, Kawakami R, Awschalom D D and Basov D N 2002 Phys. Rev. Lett. **89** 097203

[29] Kierspe W, Kohlhaas R and Gonska H 1967 Z. Angew Phys. **24** 28

[30] Dua P and Singh I 2004 J. Phys. Chem. Solids **65** 1573

[31] Dua P, Panwar S and Singh I 2006 Ind. J. Pure Appl. Phys. **44** 677

[32] Dua P, Ph. D. Thesis (unpublished)

[33] Jarrett H S, Cloud W H, Bouchard R J, Butler S R, Frederick C G and Gillson J L 1968 Phys. Rev. Lett. **21** 617

[34] Dua P and Singh I 2007 J. Appl. Phys. **101** 09G510

[35] Zhang Y and Sarma S Das 2005 Phys. Rev. Lett. **95** 256603

[36] Hirsch J E 1999 Phys. Rev. B **59** 6256

[37] Hu R, Thomas K J, Lee Y, Vogt T, Choi E S, Mitrovic V F, Hermann R P, Grandjean F, Canfield P C, Kim J W, Goldman A I and Petrovic C Cond-mat / **0801.1354**

[38]  Sarma S Das, Fabian J, Hu X and Zutic J 36 IEEE Trans. Magn. **36** 2821

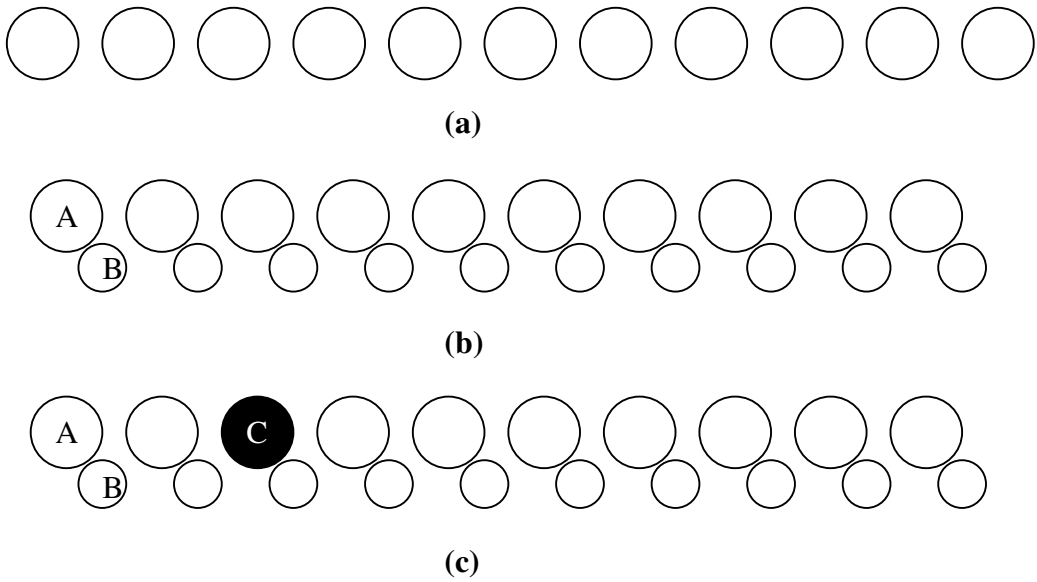

**(a)**

**(b)**

**(c)**

Fig 1. Representation of a linear chain of itinerant electron systems. A linear chain of

(a) similar atoms like Fe, Co or Ni (b) AB type system like MnAs (c) $A_{1-x}C_xB$ type system like

$Ga_{1-x}Mn_xAs$.

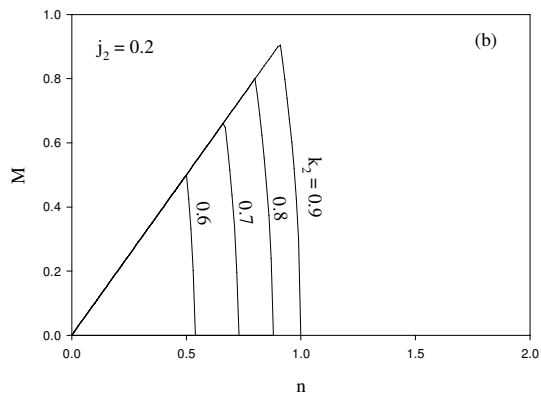

Fig 2. Saturation Magnetization as a function of band filling n

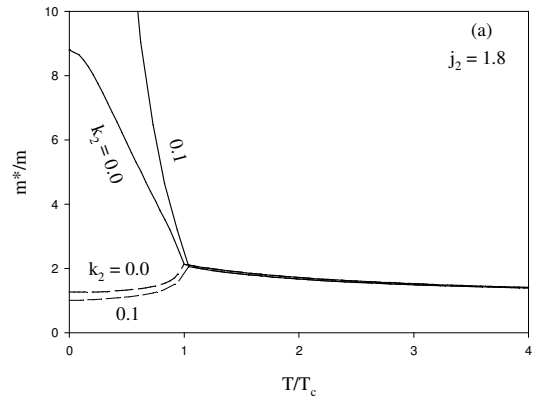

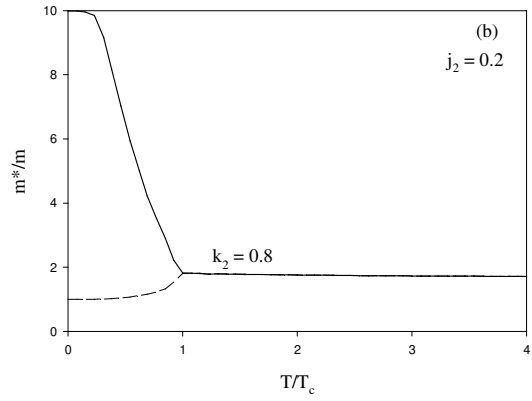

Fig 3. Effective mass ratio as a function of temperature for different values of interaction parameters

(a) $j_2 = 1.8$ (b) $j_2 = 0.2$ (-------- for up spin electrons & _______ for down spin electrons).

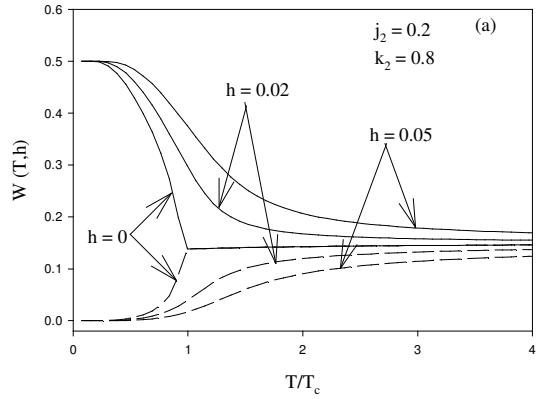

Fig 4. Spectral weight as a function of temperature for different values of interaction parameters

(________ for up spin electrons & --------- for down spin electrons).

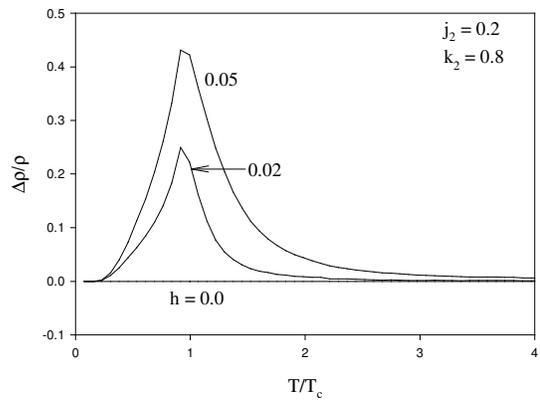

Fig 5. Magnetoresistance as a function of temperature for different values of applied magnetic

field. $j_2 = 0.2$, $k_2 = 0.8$.